# Control of spin dynamics with laser pulses: Generation of entangled states of donor-bound electrons in a $Cd_{1-x}Mn_xTe$ quantum well

J. M. Bao,[1*] A. V. Bragas,[1+] J. K. Furdyna[2] and R. Merlin[1]

[1]*FOCUS Center and Department of Physics, The University of Michigan, Ann Arbor, Michigan 48109-1120, USA*

[2]*Department of Physics, University of Notre Dame, Indiana 46556, USA*

# ABSTRACT

A quantum-mechanical many-particle system may exhibit non-local behavior in that measurements performed on one of the particles can affect a second one that is far apart. These so-called entangled states are crucial for the implementation of quantum information protocols and gates for quantum computation. Here, we use ultrafast optical pulses and coherent pump-probe techniques to create and control spin entangled states in an ensemble of up to three non-interacting electrons bound to donors in a $Cd_{1-x}Mn_xTe$ quantum well. Our method, relying on the exchange interaction between optically-excited excitons and the paramagnetic impurities, can in principle be applied to entangle an arbitrarily large number of electrons. A microscopic theory of impulsive stimulated Raman scattering and a model for multi-spin entanglement are presented. The signature of entanglement is the observation of overtones of donor spin-flips in the differential reflectivity of the probe pulse. Results are shown for resonant excitation of localized excitons below the gap, and above the gap where the signatures of entanglement are significantly enhanced. Data is also presented on the generation of coherent excitations of antiferromagnetically-coupled manganese pairs, folded acoustic phonons, exciton Zeeman beats and entanglement involving two $Mn^{2+}$ ions.

# I. INTRODUCTION

The problem of quantum entanglement has attracted much attention since the early days of quantum mechanics. For pure states, entanglement refers to non-factorizable wavefunctions that exhibit non-locality with correlations that violate Bell's inequalities.[1] As exemplified by the famous Einstein-Podolsky-Rosen paradox, one of the most intriguing features of quantum behavior is that non-interacting parts of a system can show non-local correlations reflecting interactions that occurred in the past. In this respect, entangled states of *macroscopic* systems - the so-called Schrödinger cats - are of particular interest because their properties defy classical intuition. While the quantum/classical boundary has been discussed in the theoretical literature for nearly 70 years,[1] careful experiments that probe this boundary have only been carried out in the past decade.[2] It is also recently that specific models have been solved to reveal the mechanisms by which coupling to the environment restores classical reality through decoherence.[2] Following the proposal by Deutsch for a quantum computer in 1985,[3] the building of a quantum cryptography machine in 1989,[4] and the discoveries (by Shor[5] in 1994 and by Grover[6] in 1996) of quantum algorithms that outperform those of classical computation, research on the foundations of quantum mechanics has now moved to the center of the new field of quantum information.[2] As a result, the questions of entanglement and decoherence have acquired practical significance.

The generation of a multiple-qubit entangled state is an essential step for quantum computing. Particularly important for information processing are those operations associated with gate sets that can perform any quantum computation, such as the combination of single-qubit operations with the 2-qubit controlled-NOT (C-NOT) gate which relies on entanglement.[2] Although techniques to entangle a pair of particles and, in particular, a pair of photons have been known for some time,[7] correlations involving more than two particles have been demonstrated experimentally only recently.[8,9] The various schemes that have been proposed for the implementation of a quantum computer use different strategies for attaining entangled states. These methods can be broadly categorized into two classes according to whether or not the entanglement between qubits is mediated by an auxiliary particle. In schemes based on nuclear magnetic resonance (NMR)[2] and for excitons in quantum dots,[10] the source of entanglement is the direct spin-spin exchange or Coulomb interactions which cannot be controlled experimentally. An example belonging to the second class is the ion trap system where center-of-mass phonons induce coupling between initially non-interacting ions.[9] Following specific proposals for entanglement, many systems have been considered as candidates for the implementation of a quantum computer. Schemes based on NMR, trapped ions and cavity quantum electrodynamics are just a few examples.[11,12,13] In addition, solid-state approaches (using, for example, excitons in quantum dots,[14] spins of localized electrons[15] and Josephson junctions[16]) have attracted much attention, motivated mainly by the fact that solid-phase processing techniques allow for easy integration and scaling.

In our work, the qubits are embodied by the spins of electrons bound to donors in a $Cd_{1-x}Mn_xTe$ quantum-well (QW). Our entanglement scheme, relying on an optically-excited exciton to introduce correlations between the donor-bound electrons, belongs to the second category, as described above.[17,18,19] For small $x$, $Cd_{1-x}Mn_xTe$ is a dilute magnetic semiconductor (DMS) with the zinc-blende structure. DMS are alloys for which many physical properties such as the lattice parameter and the bandgap can be widely tuned by varying the concentration of a magnetic ion.[20,21] This tunability makes $Cd_{1-x}Mn_xTe$ a very useful system for bandgap engineering and device applications. Depending on $x$ and on temperature, $Cd_{1-x}Mn_xTe$ exhibits paramagnetic, antiferromagnetic or spin-glass phases.[20,21] Many of the unique properties of DMS materials stem from the so-called *sp-d* exchange interaction involving the localized *d*-states of the magnetic ions and *sp*-states associated with the semiconductor gap. At relatively small manganese concentrations, the *sp-d* exchange leads to unusually large Zeeman splittings with concomitant giant Faraday rotation and enhancement of the electron and hole gyromagnetic factor.[20,21] Hence, the effective magnetic field experienced by states near the gap can be significantly larger than the external field. This applies also to donor-bound electrons since the corresponding wave functions derive from states near the bottom of the conduction band.[20] Due to a combination of spin-orbit coupling and quantum confinement, the heavy-hole spin in QW's made of $Cd_{1-x}Mn_xTe$, or other zinc-blende semiconductors, points along the sample growth direction, the *z*-axis, independently of the direction of the external magnetic field (this, provided the Zeeman splitting is small compared with the separation between the heavy- and light-hole states).[22,23] This property is crucial to our entanglement method for the interaction between the photoexcited heavy-hole of the exciton and the bound electrons provides the effective coupling which leads to entanglement.[24,25]

This paper is organized as follows. In Sec. II, we present a brief introduction to the theory of impulsive stimulated Raman scattering and our model for multi-spin entanglement. Experimental observations and discussions are given in Sec. III. While this paper is primarily centered on the generation of spin-entangled states of donor-bound electrons, we also report on the observation of other coherent excitations, specifically, exciton Zeeman beats in the Faraday configuration, Raman beats of antiferromagnetically-coupled manganese neighbors and folded acoustic phonons. Indirect evidence of entanglement involving at least two $Mn^{2+}$ ions is also discussed.

## II. MODEL FOR THE GENERATION OF MULTI-SPIN ENTANGLEMENT

### A. Stimulated Raman scattering and coherent superposition states

Ultrafast laser pulses can be used to generate macroscopic coherence for extended modes as well as correlations between localized quantum states through impulsive stimulated Raman scattering (ISRS).[26,27,28] In particular, ISRS has been extensively applied to generate coherent vibrations in solids and molecules (we note that the theory of ISRS by phonons, as discussed in, e. g., Ref. [28] can be easily modified to account for other modes, such as intersubband density oscillations[29] and plasmons[30]). Because of its relevance to our method for spin entanglement, we present in the following a simplified discussion of ISRS for molecular-like systems, emphasizing its main features. A more complete treatment of stimulated Raman scattering can be found in most nonlinear optics textbooks; see, e. g., Refs. [31] and [32].

The analysis of time-resolved pump-probe experiments can be conveniently divided into two parts. First, the pump pulse generates a coherent superposition state and, then, a time-delayed and weaker probe pulse is used to measure the changes in the optical constants arising from the pump-induced coherence. The relevant Hamiltonian is

$$H = H_0 + V(t) \qquad (1)$$

where the spectrum of $H_0$ is shown in Fig. 1(a). Here,

$$V(t) = -\mathbf{d}.\mathbf{F}(t) \qquad (2)$$

describes the interaction of the medium with a classical light field, $\Omega = (E_2 - E_1)/\hbar$ where $E_m$ ($m = 1,2, ..$) denotes an eigenenergy of $H_0$, $\mathbf{d}$ is the electric-dipole operator and $\mathbf{F} = \mathbf{F}_0$ ($\tilde{\mathbf{F}}_0$) is the incident time-dependent electric field of the pump (probe) pulse. To simplify the notation, $\mathbf{d}$ and $\mathbf{F}$ are treated in this section as scalars. The low-lying states |1> and |2> are associated with spin or

vibrational degrees of freedom whereas the set |l> represents higher-lying electronic states with $E_l \gg \hbar\Omega$ (in our problem, the energies of |l> are close to the QW bandgap). We assume that dipole transitions between both |1> and |2>, and the excited states |l> are allowed, i. e., $d_{li} = <l|d|i> \neq 0$ for $i = 1, 2$, and also that the system is initially in the ground state |1>. Following the interaction with the pump pulse, and ignoring decay, the wave function of the system at times that are large compared with the pulse width can be written as

$$\Psi(t) = C_1 e^{-iE_1 t/\hbar} |1> + C_2 e^{-iE_2 t/\hbar} |2> + \sum_l C_l e^{-iE_l t/\hbar} |l> \qquad (3)$$

where $C_1$, $C_2$ and $C_l$ are time-independent coefficients. To lowest order in the electric field, $C_1 \approx 1$, $C_l \approx (i/\hbar) d_{l1} F_T(\omega_l - \omega_1)$ and

$$C_2 = \int F_T(\omega) F_T^*(\omega - \Omega) \Pi(\omega) d\omega \qquad . \qquad (4)$$

Here

$$\Pi = -\frac{i}{2\pi\hbar^2} \sum_l \frac{d_{2l} d_{l1}}{[\omega - \omega_l + \omega_1]} \qquad (5)$$

and $F_T(\omega) = \int F_T(t) e^{-i\omega t} dt$ is the Fourier transform of the transmitted pump field. Note that $C_2$ vanishes if the pump bandwidth is small compared with $\Omega$. Because the expression for $\Pi$ is identical to that of the spontaneous Raman scattering matrix, the process by which the correlation between |1> and |2> is established is known as stimulated Raman scattering.[31,32] We recall that, in the spontaneous case, an incident photon of frequency ω induces a transition from the ground state |1> to some intermediate state, and it is then re-emitted as a photon of frequency ω-Ω, following the transition from the intermediate to the final state |2>. Unlike the stimulated process, for which the coherence is driven by a pair of classical fields, namely, $F_T(\omega)$ and $F_T(\omega-\Omega)$, which are contained in the pulse spectrum of Fig. 1(b), spontaneous Raman events involve single photons which leave the system in a mixed as opposed to a coherent superposition state.

In the non-resonant case, that is, for $\omega_C \ll E_l/\hbar$, where $\omega_C$ is the central frequency of the laser pulses, $C_l \approx 0$ and, provided $\Omega \ll \omega_C$, the interaction of the probe pulse with $\Psi$ can be described in terms of the slowly-varying optical susceptibility

$$\chi(\omega;t) \approx \sum_l \frac{\langle \Psi|d|l\rangle\langle l|d|\Psi\rangle}{\hbar(\omega-\omega_l)} \approx \chi_0(\omega) + C_2^* A(\omega)e^{i\Omega t} + C_2 A^*(\omega)e^{-i\Omega t} \quad (6)$$

where $\chi_0(\omega)$ is the linear susceptibility and $A(\omega) = \sum_l d_{2l}d_{l1}\hbar^{-1}(\omega-\omega_l)^{-1}$. Except for a constant factor, the expressions for $A$ and $\Pi$ are the same. As discussed in Ref. [33], this result applies strictly to transparent substances. Under resonant conditions, $A$ as well as the real part of $\Pi$ remain identical to the Raman matrix, but the imaginary component of $\Pi$ differs considerably.[33] Even though the matrix elements that apply to the generation and detection are different for opaque media, the selection rules for spontaneous and stimulated Raman scattering are the same and, as such, their determination is crucial for identifying the mechanism responsible for the coherence.

Our pump-probe experiments were performed in the reflection geometry at nearly normal incidence. Using Eq. (6) and the expression for the reflection coefficient $r = (n-1)/(n+1)$, where $n = \sqrt{1+4\pi\chi}$ is the time-dependent refractive index, we can easily find the spectrum of the reflected probe beam, $\tilde{F}_R(t)$. Writing the incident probe field as $\tilde{F}_0(t) = \int \tilde{F}_0(\omega)e^{-i\omega(t-\tau)}d\omega$ where $\tau$ is the pump-probe delay, we obtain to lowest order in both the pump and probe fields

$$\Delta\tilde{F}_R(\omega) = \tilde{F}_R(\omega) - \frac{(n_R-1)}{(n_R+1)}\tilde{F}_0(\omega) = \frac{4\pi}{n_R(n_R+1)^2}[A(\omega)e^{i\Omega\tau}\tilde{F}_0(\omega-\Omega) + A^*(\omega)e^{-i\Omega\tau}\tilde{F}_0(\omega+\Omega)] \quad (7)$$

where $n_R(\omega)$ is the standard refractive index. Thus, the change in the reflected intensity at frequency ω due to the Raman coherence is

$$\Delta R(\omega) = \left|\tilde{F}_R(\omega)\right|^2 - \left|\frac{(n_R-1)}{(n_R+1)}\tilde{F}_0(\omega)\right|^2 =$$

$$\text{Re}\{\frac{4\pi(n_R-1)}{n_R(n_R+1)^3}\tilde{F}_0(\omega)[A^*(\omega)\tilde{F}_0^*(\omega-\Omega)e^{-i\Omega\tau} + A(\omega)\tilde{F}_0^*(\omega+\Omega)e^{i\Omega\tau}]\} \quad (8)$$

which can be written as

$$\Delta R \equiv \gamma(\omega)\sin[\Omega\tau + \phi(\omega)] \quad . \quad (9)$$

It follows that the reflectivity oscillates with frequency Ω as a function of the time delay τ. Similar results are obtained for the phase of the reflection coefficient and for the magnitude and phase of the transmission coefficient. However, in the transmission geometry the dominant term is proportional to the length of the sample and has an additional phase of π/2.[27,28] Analogous to Eq. (4), Eq. (8) indicates that the differential reflectivity depends on the overlap between $\tilde{F}_0(\omega)$ and $\tilde{F}_0(\omega-\Omega)$. Also, note that $\Delta R$ is proportional to the product of the pump and probe intensities [27,28]. Finally, it can be shown for multiple coherences (more than two correlated low-lying levels) that the total signal is a linear superposition of terms such as those in Eq. (9) with different frequencies.

Correlations involving low-lying levels are known as ground-state coherences.[34,35] For resonant excitation and, in particular, when the pulse width is greater than the energy separation between states in the |l> - manifold, the pump pulse can also induce correlations between these higher-lying levels which are referred to as excited-state coherences. Like ground-state coherences, excited-state coherences also modify the reflected probe spectrum (although not by way of Raman but through stimulated emission)[34,35] and these changes can be measured in standard pump-probe experiments.

**B. Multi-spin coherence: donors and Mn$^{2+}$ spins in Cd$_{1-x}$Mn$_x$Te**

For two or more particles to be entangled, it is generally required that they be interacting or that an interaction had occurred in the past. In our case, an optically excited exciton provides the coupling between $N$ electrons bound either to donors or to Mn$^{2+}$ ions. The relevant Hamiltonian is

$$H = H_S + V(t) = |g>H_g<g| + |e>H_e<e| + V(t) \quad (10)$$

where $|g>$ is the ground state of the solid and $|e>$ represents an exciton of energy $E_e$.[22,36] As before, $V(t) = -\mathbf{d}\cdot\mathbf{F}(t)$ accounts for the interaction with light pulses. $H_g$ describes the Zeeman coupling of the electrons with the external magnetic field, given by

$$H_g = \sum_{i=1}^{N} g\mu_B \mathbf{s}_i \cdot \mathbf{B} = g\mu_B \mathbf{S}\cdot\mathbf{B}, \quad (11)$$

where $\mathbf{s}_i$ is the spin of the $i$th electron, $\mathbf{S} = \sum_{i=1,..,N} \mathbf{s}_i$ is the total spin of $N$ electrons, $\mathbf{B}$ is the applied magnetic field and $g$ is the appropriate gyromagnetic factor. For the $d$-states of Mn$^{2+}$, $g_{Mn} \approx 2$ whereas, thanks to the presence of manganese, the gyromagnetic factor of both, electrons and holes in Cd$_{1-x}$Mn$_x$Te is considerably larger than in CdTe (see Sec. III).[20] The term

$$H_e = E_e + g\mu_B \mathbf{S}\cdot\mathbf{B}_w \quad (12)$$

describes the interaction of the bound electrons with the exciton. Here $\mathbf{B}_w = \mathbf{B} + \mathbf{B}_e$ is the effective magnetic field, $\mathbf{B}_e = -\kappa\mathbf{J}/g\mu_B$ is the exchange field where $\mathbf{J}$ is the spin of the heavy-hole composing the exciton ($|\mathbf{J}| = 3/2$) and $\kappa$ is a coupling constant describing the exchange interaction between electrons and heavy-holes, which depends on the overlap between their wave functions (for simplicity, we assume that κ is the same for all donor electrons).[24,25,37,38,39] As mentioned earlier, provided the external magnetic field is not too strong, **J** is parallel to the *z*–axis.[23,40]

The Hamiltonian $H_S$ in Eq. (10) can be diagonalized exactly. The eigenfunctions are separable into products of bound-electron spin and exciton states. The energy levels are shown schematically in Fig. 2. The states at the bottom of the figure are of the form $|S-k\rangle_x \otimes |g\rangle$, where $|S-k\rangle_x$ is one of the Zeeman-split states of the total spin **S** with the quantization axis parallel to the *x*-direction. Here, $\mathbf{S}^2|S-k\rangle_x = S(S+1)|S-k\rangle_x$ and $S=N/2$. Also, $H_S|S-k\rangle_x \otimes |g\rangle = k\hbar\Omega_0 |S-k\rangle_x \otimes |g\rangle$ with $0 \leq k \leq 2S$ where $\Omega_0 = g\mu_B B/\hbar$ is the paramagnetic resonance (PR) frequency associated with the spin–flip of a single electron. In the presence of the exciton, $H_S|S-l\rangle_w \otimes |e\rangle = (E_e + l\hbar\Omega'_0)|S-l\rangle_w \otimes |e\rangle$ where $|S-l\rangle_w$ is a Zeeman-split state for which the quantization axis is parallel to $\mathbf{B}_w$ and $\Omega'_0 = g\mu_B B_w/\hbar$. As mentioned in Sec. I, the fact that **J** is parallel to the *z*-axis or, alternatively, that $\mathbf{B}_w$ and **B** are in different directions, is crucial to the implementation of our method. If $\langle g|\mathbf{d}|e\rangle \neq 0$, transitions between $|S-k\rangle_x \otimes |g\rangle$ and $|S-l\rangle_w \otimes |e\rangle$ are electric-dipole allowed because Zeeman states along different quantization axes are not orthogonal to each other. Hence, a Raman coherence between different states in the $|S-k\rangle_x \otimes |g\rangle$ manifold can be attained by using two dipole-allowed transitions with $|S-l\rangle_w \otimes |e\rangle$ as intermediate states (note that, since $\langle S|S'\rangle = \delta_{SS'}$, states with $S \neq N/2$ are not accessible to our method). As an example, the two arrows in Fig. 2 denote transitions which can lead to the so-called maximally entangled Bell state $(|-S\rangle_x + |S\rangle_x)/\sqrt{2}$. More generally, a properly tailored optical pulse can, in principle, generate a predefined coherent superposition state of the form

$$\psi(t) = \sum_{k=0}^{2S} C_k e^{-ik\Omega_0 t} |S-k\rangle_x \qquad . \tag{13}$$

For $S \gg 1$, our problem can be mapped into that of two harmonic oscillators displaced with respect to each other as represented by the parabolas in Fig. 2. Explicitly

$$\lim_{S \to \infty} H_e = E_e - g\mu_B B(S+\frac{1}{2}) - \frac{S\kappa^2 J_z^2}{2\mu_B gB} + \frac{g\mu_B B}{2S}(S_z - \Delta S)^2 \tag{14}$$

where $J_z = \pm 3/2$ and $\Delta S = S\kappa J_z / \mu_B gB$ is the spin displacement.[41] Thus, the Hamiltonian becomes identical to that of a molecule with electrons that couple to a single vibrational mode. This mapping, generalized to incorporate anharmonicity (to account for a finite number of spin levels), is important for it suggests that spin manipulation can be attained by using the same coherent-control techniques that apply to molecular vibrations. Available techniques include pulse shaping, four-wave-mixing, stimulated Raman adiabatic passage (STIRAP) and pump-dump methods.[42,43,44,45] The displaced-oscillator model also provides a natural explanation for the occurrence of vibrational overtones in spontaneous Raman scattering[36,46] and, relevant to our work, it has been used to account for Raman observations of up to 15 overtones of the $Mn^{2+}$ spin-flip PR-transition in a $Cd_{1-x}Mn_xTe$ QW.[22]

Based on the discussion of Sec. IIA, the coherent superposition state of Eq. (13) is expected to modulate the optical constants of the sample at frequencies that are multiples of $\Omega_0$ (in comparison, the modulation due to an arbitrary non-entangled product state should only contain the fundamental frequency). Because the electron spin is $s = \frac{1}{2}$, the presence of the $m$th harmonic of $\Omega_0$ is the signature of an entanglement involving $m$ electrons. However, it is important to call attention to the fact that we cannot distinguish between a system of non-entangled electrons and entangled states in the $S \to \infty$ limit since the impulsive excitation of a strictly harmonic oscillator creates a Glauber coherent state which does not exhibit overtones.[28] In our ISRS experiments, we detect the first and second overtone of the PR for donor-bound electrons, but only the fundamental mode for $Mn^{2+}$ electrons. These results are consistent with the displaced-oscillator model in that a system of two or three donors is considerably more anharmonic than the five-electrons of a single $Mn^{2+}$ ion.

<h3 style="text-align:center">III. EXPERIMENTS</h3>

Our sample is a 100-period superlattice consisting of, nominally, 58-Å-thick CdTe wells with 19-Å-thick MnTe barriers, grown by molecular-beam-epitaxy on a thick (relaxed) [001] CdTe substrate. Bulk CdTe is a non-magnetic zinc-blende semiconductor with a bandgap of 1.6eV and MnTe is an antiferromagnet with gap at 3.23 eV.[20,21] Due to diffusion from the barriers, the wells in our sample contain a small number of $Mn^{2+}$ ions close to the interfaces so that, instead of CdTe, we have, on average, $CdTe_{0.996}Mn_{0.004}Te$. Although the amount is minor, the presence of manganese impurities has a profound effect on the magnetic response of the QW. The superlattice is nominally undoped. Consistent with other reports,[47,48,49] however, our Raman experiments reveal the presence of isolated donors in the wells (possibly indium) with a concentration of $\sim 5\times10^{16}$ cm$^{-3}$.

We used the 488.0 nm line of an $Ar^+$-ion laser to obtain photoluminescence (PL) data, and a home-built continuous-wave tunable Ti-sapphire laser operated at a power density of $\sim 10^{-2}$ Wcm$^{-2}$ to acquire spontaneous Raman scattering and photoluminescence excitation (PLE) spectra. Our split-coil superconducting optical cryostat provides fields up to 7 T.

Differential reflectivity data was obtained using a standard pump-probe setup. As light source, we used a mode-locked Ti-sapphire laser, pumped by a 532 nm, 5 W solid state laser which provided ~ 100 fs pulses at the repetition rate of 82 MHz which were focused on a 400-µm-diameter spot using an average power of 3-4 mW. The central wavelength of the pulses could be tuned in the range 720-776 nm to resonate with the QW bandgap. Data were obtained either in the Voigt ($\mathbf{B} \perp z$) or the Faraday ($\mathbf{B}//z$) geometry with the photon wavevector along the $z$-axis. Unless stated otherwise, the pump beam was circularly polarized to couple to a single spin component of the heavy-hole (say, $J_z = 3/2$) while the incident probe beam was linearly polarized. We measured the pump-induced change in the field of the reflected probe beam, $\Delta \tilde{\mathbf{F}}_R$, as a function of the time delay between the two pulses; see Eq. (7). To obtain $\Delta \tilde{\mathbf{F}}_R$, we determined separately the pump-induced shift of the polarization angle of the reflected probe field, $\Delta\theta$, and the differential reflectivity $\Delta R \propto \tilde{\mathbf{F}}_R \cdot \Delta \tilde{\mathbf{F}}_R$; see Eq. (8). We worked in the linear regime in which both $\Delta\theta$ and $\Delta R$ are proportional to the intensity of the laser pulses. $\Delta\theta$ was gained from magnetic Kerr measurements using the scheme described in Refs. [50,51] which gives an output signal proportional to $\tilde{\mathbf{F}}_R \times \Delta \tilde{\mathbf{F}}_R$. A symmetry analysis indicates that $\Delta \tilde{\mathbf{F}}_R$ must be perpendicular (parallel) to $\tilde{\mathbf{F}}_R$ and, hence, that $\Delta R \equiv 0$ ($\Delta\theta \equiv 0$) for excitations involving an odd (even) number of spin-flips. Such excitations are odd (even) under time-reversal and, thus, they belong to the antisymmetric $A_2$ (symmetric $A_1$) representation of the $D_{2d}$ point group of the QW.

### A. Sample characterization: Photoluminescence and spontaneous Raman scattering

#### 1. PL and PLE

Figure 3 shows the PL and PLE spectra of our sample. Features below 1.71 eV are associated with transitions involving heavy-hole states. Due to quantum confinement, the QW bandgap is blue shifted, by ~ 80 meV, with respect to the bandgap of bulk CdTe. Consistent with other studies on similar samples,[22] the PL peak is red-shifted (by ~ 6 meV) with respect to the first

PLE maximum, an indication that the PL is associated with the recombination of localized excitons. The PLE feature at ~ 1.73 eV is attributed to transitions involving the light-hole subband, as supported by a calculation of the QW level structure. Such transitions are outside the range of our Raman and pump-probe experiments.

In DMS systems, the interaction of the electron and hole with the magnetic ions leads to an unusual magnetic-field behavior of the exciton energy.[20,21] Schematic diagrams illustrating the field dependence of conduction and heavy-hole states as well as the selection rules for excitation with circularly-polarized light are shown in Fig. 4. For conduction-band states in $Cd_{1-x}Mn_xTe$, the Zeeman coupling gives[20,21]

$$E^c_{\pm} = \pm(\frac{1}{2}g_C\mu_B B - N_0\alpha x <S_B>/2) \tag{15}$$

where the quantization axis is along the field direction. Here $E_+$ and $E_-$ are the energies of the states with spin up and down, $g_C = -1.6$ is the bare electron gyromagnetic factor in CdTe, $N_0$ is the number of cations per unit volume, $N_0\alpha \approx 220$ meV characterizes the *sp-d* interaction for the electron, $x$ is the concentration of $Mn^{2+}$ ions in the CdTe well and $S_B$ is the component of the spin of a single $Mn^{2+}$ ion parallel to the magnetic field. Hence, $\langle S_B \rangle = -\frac{5}{2}B_{5/2}(\mu_B gB/k_B T)$ where $B_{5/2}$ is the Brillouin function for $S = 5/2$. An expression similar to Eq. (15) applies to the heavy-hole in the Faraday configuration, namely,[20]

$$E^{HH}_{\pm} = \pm(\frac{3}{2}g_{HH}\mu_B B - N_0\beta x <S_z>/2) \tag{16}$$

where $N_0\beta \approx -880$ meV. For the gyromagnetic factor of the hole in a CdTe QW, the approximate value $g_{HH} \approx 0.65$ can be gained from results reported in Ref. [52]. Once again, we emphasize the fact that, at small fields, the heavy hole exhibits no Zeeman splitting in the Voigt configuration since its spin is oriented along the $z$-axis.[23]

The measured magnetic-field dependence of the PL is reproduced in Fig. 5(a). Fig. 5(b) shows the strongly nonlinear dependence of the PL red-shift which saturates at large fields, closely following the behavior of the Brillouin function. The curves in Fig. 5(b) are fits to ($E_-^C + E_-^{HH}$) for the Faraday and $E_-^C$ for the Voigt configuration. From these fits, the manganese concentration can be determined. The data in the Voigt (Faraday) configuration gives $x = 0.004$ (0.0035) and $T = 3.5$ K. The fact that the temperature from the fit is slightly higher than the bath temperature is attributed to laser heating. We believe that the manganese concentration obtained from measurements in the Voigt configuration is closer to the actual value for our sample because of the uncertainty in the determination of $g_{HH}$, which exhibits a strong dependence on well-width.[52] Moreover, as shown below, fits to spontaneous Raman and pump-probe data on the magnetic-field dependence of the donor spin-flip transition also give $x = 0.004$.

## 2. Resonant spin-flip Raman scattering

Raman scattering has been widely applied to study electronic excitations of impurities in a broad range of materials.[53] In particular, spin-flip transitions of magnetic ions and donor-bound electrons have been investigated at length in DMS (both bulk and QW form).[47,48] Raman spectra obtained in the Voigt geometry using laser energies near the QW bandgap are shown in Figs. 6 and 7. The incident (scattered) light is polarized along the [110] ([1$\bar{1}$0]) axis. Most of the features are due to transitions within the Mn$^{2+}$ $^6S_{5/2}$ - multiplet. The spectrum of Fig. 6 shows the manganese PR (i. e., the transition $S_z = 5/2 \rightarrow S_z = 3/2$), two PR overtones and, in addition, the peak labeled 1SF at ~ 12 cm$^{-1}$ due to the spin-flip of an electron bound to a donor.[17,22,49] The PR frequency shifts linearly with $B$ with a slope consistent with $g_{Mn} \approx 2$.[47] As shown in the inset of Fig. 6, the 1SF behavior follows closely the theoretical prediction, Eq. (15), in that its frequency is approximately twice the frequency shift of the PL in the Voigt configuration (we note that $g$ has approximately the same value for free electrons, electrons bound to donors and electrons in free and bound excitons). The 1SF width is ~ 3 cm$^{-1}$. We believe that the broadening is primarily due to fluctuations in the local Mn$^{2+}$ concentration.

As shown in Fig. 7, the manganese overtone scattering is strongly enhanced when the laser energy is tuned to resonate with the PL feature at ~ 1.68 eV (see Fig. 3). The PR intensity versus laser excitation energy is plotted in the inset of Fig. 7. The blue-shift of the Raman maximum with respect to the PL peak indicates that the scattering is stronger for the outgoing resonance and, together with the results of Fig. 3, that localized excitons are the relevant intermediate states in the multiple spin-flip Raman process.[22] Such a behavior resembles what was reported many years ago for donor spin-flip harmonics in CdS and ZnTe.[54,55,56] Within the context of the model of Fig. 2, the observation of more than five overtones is an indication that a single exciton couples on average to at least two Mn$^{2+}$ ions. While the donor spin-flip also resonates with the localized excitons, the stronger manganese scattering prevented us from making an accurate determination of its resonant behavior. We further note that the spectra reveal no evidence of 1SF overtones. Therefore, spontaneous Raman data cannot be used to determine the average number of donors that couple to a single exciton (as shown in the next section, the time-domain measurements give three donors per exciton).

Our Raman results are in excellent agreement with the work of Stühler *et al.* on similar samples.[22] Consistent with the model depicted in Fig. 2, the multiple manganese PR-scattering is not observed in the Faraday geometry where **B**, **B**$_e$ and the hole quantization axis are all along *z*.

Moreover, peaks involving an odd (even) number of $Mn^{2+}$ spin-flips are seen mainly when the polarizations of the incident and scattered light are perpendicular (parallel) to each other, reflecting the fact that the corresponding excitations belong to the $A_2$ ($A_1$) representation of the $D_{2d}$ point group. These selection rules apply at temperatures that are not too low. As shown in Fig. 7, a departure from these rules is observed at low temperatures suggesting that the actual symmetry of the QW is lower than $D_{2d}$; see Ref. [22] and, in particular, Ref. [49] for related observations on 1SF.

### B. Ultrafast pump-probe experiments and entanglement generation

The time-domain results discussed below reveal spin-related oscillations associated with the same transitions observed in Raman spectra and, in addition, new features identified as excited-state coherences. The amplitude of the coherent oscillations shows a strong dependence on both the magnetic field and the central energy of the pulses and, consistent with Raman results, the selection rules show departures from the nominally $D_{2d}$ symmetry of the QW. An important distinction with the frequency-domain data is that, while the Raman spectra of $Mn^{2+}$ exhibits pronounced PR-overtones, the time-domain traces show only the fundamental frequency. In some sense, the opposite applies to electrons bound to donors. Unlike spontaneous Raman data, for which the corresponding spin-flip overtones are too weak to be observed, up to two overtones are detected in the pump-probe experiments. As discussed in Section IIB, this observation is the signature of a three-electron entanglement.

#### *1. Pump-probe oscillations: laser tuned below the QW bandgap*

Pump-probe data with the laser tuned below the QW bandgap is shown in Fig. 8. The oscillations were analyzed using linear prediction (LP) methods,[57] and the parameters from the LP fit were used to generate the Fourier transform spectra shown in the insets. The top trace in Fig. 8 shows a single oscillator of asymmetric lineshape at ~ 5.3 cm$^{-1}$. The frequency of this mode varies linearly with the magnetic field with a slope that agrees extremely well with the electron gyromagnetic factor of bulk CdTe. Consequently, these oscillations are assigned to the spin precession of electrons in the CdTe substrate. A somehow broader remnant of this signal is also visible in the bottom spectrum, obtained at a slightly higher central frequency (as shown below, the substrate signal disappears as we tune the laser closer to the QW bandgap). In addition, the results of Fig. 8(b) show short- and long-lived oscillations of frequency ~ 9 cm$^{-1}$ and ~ 6.5 cm$^{-1}$ which dominate, respectively, during the first ~ 20 ps and above ~ 30 ps. The long-lived feature is the manganese PR.[51] The measured gyromagnetic factor, $g \approx 2$, agrees extremely well with values from Mn$^{2+}$ spin-flip Raman scattering (Figs. 6 and 7)[47] and other experiments.[48,51] The broad feature at ~ 9 cm$^{-1}$ is assigned to the spin-flip of donor electrons. Its width and the magnetic-field dependence of its frequency are nearly the same as that of the 1SF line in Fig. 6 (note that the frequency from time-domain is slightly lower than that obtained from Raman data because the pump-probe experiment was performed at a higher temperature).

### *2. Pump-probe oscillations: laser in resonance with localized excitons*

Figure 9(a) shows differential magnetic-Kerr data for excitation resonant with localized excitons. These results have a close similarity to those reported for (Zn,Cd,Mn)Se heterostructures by Crooker et al. who ascribed the rapidly decaying oscillations to free photoexcited electrons.[51] However, close inspection of the data brings out important differences concerning the nature of the oscillations as well as the source of the coherence. Other than the oscillations associated with the spin-flip of electrons bound to Mn$^{2+}$, at ~ 5.5 cm$^{-1}$, the LP fit reveals peaks at ~ 7, 11 and 12 cm$^{-1}$ as well as a weaker feature at ~ 22 cm$^{-1}$ which is the first overtone of the 11 cm$^{-1}$ peak; see later. The dependence of the frequency of these oscillations on magnetic field is plotted in Fig. 9(b). As before, PR denotes the manganese paramagnetic resonance showing a long decay time, whereas the feature labeled PR(e), at a slightly higher frequency, is attributed to its excited-state counterpart, that is, the manganese spin-flip in the presence of the exciton.[17] The doublet identified as SF follows closely the expected behavior of the spin-flip transition of a conduction electron. Accordingly, and for other reasons discussed below, its low- and high-frequency components are assigned, respectively, to the donor spin-flip ground-state coherence (the Raman counterpart in Fig. 6 is 1SF) and the electron spin-flip of photoexcited excitons[17] (or, alternatively, exciton quantum beats involving the two optically-allowed transitions in the Voigt configuration; see Fig. 4). The fact that the electron spin-flip frequency decreases slightly with $B$ at high fields, and a somewhat different operating temperature, explain why the 1SF-frequency in Fig. 9(a) is higher than in Fig. 8(b). Results using the same laser pulses as in Fig. 9(a), but smaller magnetic fields, are shown in Fig. 10. The Fourier-transform insets show only the lower component of the SF-doublet, indicated by 1SF, and its second harmonic, 2SF. Additional differential magnetic Kerr traces obtained using resonant excitation of localized excitons are shown in Fig. 11(a). Besides the first and second harmonic of the donor spin-flip excitation, the LP-fits for $B = 3.4$ and $6.7$ T reveal the third harmonic, 3SF. The field dependence of the multiple spin-flip frequencies is plotted in Fig. 11(b).

Our assignment of PR(e) as due to the excited-state manganese PR is supported by the following argument. Consider the heavy-hole component of the exchange interaction $V_{HH} = (\beta/3)\sum_i \delta(\mathbf{r}-\mathbf{r}_i)\mathbf{S}_i \cdot \mathbf{J}$ between an exciton and $Mn^{2+}$ ions at sites $\mathbf{r}_i$ where β is the constant defined in Eq. (16) and $\mathbf{r}$ is the heavy-hole position.[20,58] As mentioned in Sec. IIB, the effect of $V_{HH}$ on the ions can be expressed in terms of an effective field $\mathbf{B}_e(\mathbf{r}) = (\beta/3\mu_B g_{Mn})|\Psi_{HH}(\mathbf{r})|^2 \mathbf{J}$ where $\Psi_{HH}$ is the hole wavefunction.[22] Hence, in the presence of the exciton the paramagnetic

transition energy should evolve from $\mu_B g_{Mn} B_e$ at $B = 0$ to $\mu_B g_{Mn} B$ for $B \gg B_e$. This prediction is consistent with the observed behavior of PR(e) in Fig. 9(b). We believe that exchange-interaction fluctuations is the reason why the width of the PR(e) line is much larger than that of PR. From the PR(e) frequency at zero field and using $N_0\beta \approx 0.88$ eV for CdTe,[20] we obtain the very reasonable estimate of ~ 40 Å for the hole localization length (the bulk exciton radius in CdTe is 50Å) leading on average to 2.5 $Mn^{2+}$ ions per hole. From this number, and in terms of the displaced-oscillator model of Fig. 2, we infer the value ~ 1.06 for the Huang-Rhys factor at $B = 7$ T.[59] This value is consistent with the observation of ~ 10 PR overtones in the Raman spectrum. Albeit indirect, the combination of the RS and coherent results gives compelling evidence for exciton-mediated entanglement involving at least two $Mn^{2+}$ ions. Further support for our interpretation is provided by the comparison between Raman scattering and time-domain data on the laser-energy dependence of, respectively, the $Mn^{2+}$ PR intensity and $\Delta\theta$, shown in Fig. 12. Other than for the dip at the PL maximum due to absorption, and the fact that the resonance width is larger for $\Delta\theta$, due to broadening introduced by the pulse bandwidth, the resonant behavior of $\Delta\theta$ is close to that of the Raman PR-peak. We construe this as evidence that the mechanism for PR-oscillations is stimulated Raman scattering.

Our assignment of the SF doublet, 2SF and 3SF as due to the spin-flips of bound electrons is supported, first, by the observation that these features resonate at $\hbar\omega_C \approx E_e$ (not shown) and, as mentioned earlier, by their dependence on temperature and field which show excellent agreement with theoretical predictions. The curves in Fig. 9(b) are fits using the Brillouin function to account for $\langle S_B \rangle$ in Eq. (15). The resonant behavior is a clear indication that the relevant intermediate states are localized. This, and the linear dependence of the signal with the pump intensity are consistent with the donor interpretation for 2SF and 3SF since linearity excludes the possibility that the overtones could be due to multiple spin-flips of bound excitons. The fact that 2SF is much weaker than 3SF is also consistent with our assignment since double-flip excitations transform like $A_1$ and, therefore, are nominally forbidden in the geometry we used (modes associated with an odd number of flips belong to the $A_2$ representation, and they are allowed). Based on the value of the frequencies at large fields, the fundamental mode from which 2SF and 3SF derive is ascribed to the lower-frequency component of the SF doublet. This mode and the overtones exhibit a similar $\hbar\omega_C$–behavior, different from that of the higher-frequency component associated primarily with the spin-flip of the electron in the bound exciton (see diagram in Fig. 4). The fact that the energy of the latter is slightly larger is attributed to exchange effects. From the splitting, we obtain an upper limit of ~ 600 μeV for the electron-hole exchange that is consistent with the value 270 μeV from the literature.[60] This estimate ignores electron-electron exchange which is large for exciton-donor complexes,[24] and may provide an additional longer-range mechanism for donor entanglement.

To provide a quantitative estimate of the donor entanglement, we consider *all* sets of $m$ impurities which interact with the ensemble of photoexcited excitons. Let $I_0$ be the integrated intensity of the pump pulse, and assume $T = 0$. Then, if $|0\rangle$ is the wavefunction immediately before the pulse strikes (the multi-impurity ground state with all spins aligned along **B**), integration of Schrödinger equation gives, to lowest order in $I_0$, the coherent superposition state

$$\Phi \approx |0\rangle + iI_0 \sum_{2S \geq k > 0, \eta} \Xi_{\eta k} e^{-ik\Omega_0 t} |\eta, S - k\rangle \qquad (17)$$

which is of the same form as that in Eq. (13). Here $\eta$ denotes a specific impurity set, $S = m/2$, $\Xi_{\eta k} = (4\pi / n_R c\hbar) \sum_{\alpha\beta} e_\alpha e_\beta R^\eta_{\alpha\beta}(k)$, $n_R$ is the refractive index, $\mathbf{e} = \mathbf{F}_0/F_0$ is a unit vector (as before, $\mathbf{F}_0$ is the electric field of the pump pulse) and $R^\eta_{\alpha\beta}(k)$ is the Raman tensor for the transition $|\eta, S\rangle \to |\eta, S-k\rangle$ of the particular set (this tensor vanishes unless all the impurities in the set interact with a single exciton). As discussed in Section IIA, a secondary effect of the interaction of the spin system with light is that $\Phi$ leads to time-varying optical constants with concomitant oscillations in the intensity or polarization of the reflected probe pulse, with amplitude proportional to $I_0 \sum_{\eta k} |\Xi_{\eta k}|^2 \exp(-ik\Omega_0 t)$. For a given impurity set, the only non-zero components of the density matrix operator $\hat{\rho}_\eta(k,l) = |\eta, S-k\rangle\langle\eta, S-l|$ are $\langle\Phi|\hat{\rho}_\eta(0,0)|\Phi\rangle \approx 1$ and $\langle\Phi|\hat{\rho}_\eta(k,0)|\Phi\rangle = \langle\Phi|\hat{\rho}_\eta(0,k)|\Phi\rangle^* = -iI_0 \Xi_{\eta k} \exp(-ik\Omega_0 t)$. Let $N_m$ be the total number of sets of $m$ impurities. Defining the ensemble average $\rho(m) = \sqrt{\sum_\eta |\langle\Phi|\hat{\rho}_\eta(m,0)|\Phi\rangle|^2 / N_m}$, it follows that the amplitude of the $m$th-harmonic oscillations gives a direct measure of the entanglement since it is proportional to $|\rho(m)|^2$. From the results for donors at 3.4 T in Fig. 11(a) (see also Fig. 3 of Ref. [17]), we get $\langle\rho(2)\rangle/\langle\rho(1)\rangle \approx 0.014$ and $\langle\rho(3)\rangle/\langle\rho(1)\rangle \approx 0.46$. It is important to realize that the strength of the $m$th harmonic also measures the probability of finding $m$ donors in the region where the coupling between the localized exciton and the impurities is significant. Assuming an interaction length of $\sim 100$ Å (in CdTe, the donor radius is $\sim 50$ Å), the ratio between the frequency-integrated intensities of the first (1SF) and third harmonic (3SF) gives the crude estimate of $5 \times 10^{16}$ cm$^{-3}$ for the density of donors in our sample. Using this density, we obtain $\langle\rho(1)\rangle/I_0 \approx 30$ m$^2$/J which is $\sim 10^3$-$10^4$ larger than what a calculation gives for off-resonance excitation.

### 3. Pump-probe oscillations: laser in resonance with free excitons

Like the manganese PR (see Fig. 12), the amplitude of the donor-related oscillations decreases as the central energy of the pulses moves away from the resonance with localized excitons at ~ 1.68 eV. Yet, these oscillations reappear when the laser energy is tuned to resonate with free excitons. As shown in Fig. 13 for $\hbar\omega_C = 1.70$eV and in Fig. 14 for $\hbar\omega_C = 1.71$eV, the amplitude of the second harmonic becomes comparable to that of 1SF.[18] The LP fits also show the 3SF mode but, relative to 1SF, its strength is comparable to values obtained at the localized-exciton resonance. These results clearly indicate that the two-electron entanglement benefits from the mediation of free excitons and, consistent with the Raman results for $Mn^{2+}$ spin flips at low temperatures,[22,49] that the QW symmetry is lower than $D_{2d}$ since, otherwise, $A_1$–symmetry modes such as 2SF should not exhibit magnetic Kerr oscillations. Even though the ideal QW symmetry is not expected to hold in the presence of alloy disorder and interface roughness, we have not been able to identify the particular symmetry-breaking process responsible for the two-particle entanglement. We note that a mechanism for entanglement involving the RKKY interaction between localized electrons and extended excitons has been recently proposed.[19]

### C. Exciton Zeeman beats

As discussed in Sec. IIIB2, other than spin-flip oscillations of paramagnetic impurities the experiments in the Voigt geometry reveal quantum beats associated with the electron Zeeman-split levels of localized excitons (Fig. 9). Similar to early reports for $Al_xGa_{1-x}As$-based heterostructures,[61] Fig. 15 shows in the Faraday configuration (**B**$//z$) quantum beats involving the heavy-hole exciton states $|\sigma^+\rangle = |J_z = +3/2, S_z = +1/2\rangle$ and $|\sigma^-\rangle = |J_z = -3/2, S_z = -1/2\rangle$; see the level diagram in Fig. 4. This assignment is based on the results of Fig. 16 which show that the frequency of the oscillations is very close to twice the Faraday PL shift (see Fig. 5). The beat frequency at 7 T is ~ 80 cm$^{-1}$ and its lifetime is ~ 1 ps. As for other excitations described in this work, we believe that the dominant source of beat decay is inhomogeneous broadening due to fluctuations in the manganese concentration.

The differential-reflectivity data in Fig. 15 was acquired using linearly-polarized pump and probe beams. In this configuration, the *localized* excitons created by the pump pulses are in a coherent superposition of states $|\sigma^+\rangle$ and $|\sigma^-\rangle$. Surprisingly, magnetic-Kerr data obtained with circularly-polarized pump pulses (the same setup we used to measure $\Delta\theta$ in the Voigt configuration) gives nearly identical results. This is unexpected because, for **B**//*z*, circularly-polarized light couples only to one of the two states, and also because the rotation of the probe polarization is inconsistent with the quantization axis being along the *z*-direction (light polarized normal to *z* cannot couple to the operators $S_x$ and $S_y$ which lead to spin precession). Whereas the specific reason why the Faraday quantum beats are observed in the magnetic-Kerr geometry is not known at this time, these observations add weight to the reported evidence that the symmetry of $Cd_{1-x}Mn_xTe$-based QW's is not $D_{2d}$.[22,49]

### D. Folded acoustic phonons and antiferromagnetically-coupled Mn$^{2+}$ pairs

The differential reflectivity (although not magnetic-Kerr) data reveal also oscillations for which the frequency does not depend on the magnetic field. As shown in Fig. 17, these oscillations are observed in the Faraday configuration after ~ 3 ps for all values of *B*. They can also be seen in the Voigt configuration at small fields. The Fourier spectra in the insets show two peaks, one labeled AFMR (antiferromagnetic resonance) at 6.8 cm$^{-1}$ and the second one at 11.3 cm$^{-1}$. These modes have entirely different origins. The higher-frequency one is due to coherent acoustic phonons of the $CdTe_{0.996}Mn_{0.004}Te$-MnTe superlattice which propagate along the *z*-axis, while AFMR is a transition associated with a pair of antiferromagnetically-coupled Mn$^{2+}$ nearest neighbors.[62,63,64] It is important to observe that the latter is unlike the problem discussed in previous sections where the manganese ions that become entangled are several lattice constants apart and their coupling is mediated by excitons.

Our assignment of the AFMR-peak is based on a comparison with existing spontaneous[62,63] and stimulated Raman data[64] on similar DMS samples. For **B** // *z*, we write the Hamiltonian describing the interaction between neighboring $Mn^{2+}$ ions of spin $\mathbf{s}_1$ and $\mathbf{s}_2$ as

$$H_{AF} = -2J_{AF}\mathbf{s}_1 \cdot \mathbf{s}_2 + g_{Mn}\mu_B B(s_{1z} + s_{2z}) = -J_{AF}[S(S+1) - \frac{35}{2}] + g_{Mn}\mu_B B S_z \quad (18)$$

where $\mathbf{S} = \mathbf{s}_1 + \mathbf{s}_2$ ($S = 0, \ldots, 5$ and $S_z = -S, \ldots, +S$) and $J_{AF}$ is the antiferromagnetic coupling constant. Measured values of $J_{AF}$ for nearest neighbors are in the range 3–4.2 $cm^{-1}$.[62,63,64] The position of AFMR agrees extremely well with values reported for the dominant Raman transition between the ground state of the pair, with $S = 0$, and the $S_z = 0$ state of the $S = 1$ triplet, which, consistent with the fact that AFMR is only observed in differential-reflectivity measurements, is even under time reversal.[62,63,64] We note that the energy of this transition is given by $2J_{AF}$ and does not depend on *B*.

High frequency ($\geq 0.1$ THz) acoustic phonons are usually not accessible to light scattering experiments because of momentum mismatch. In artificial periodic structures such as superlattices, however, modes at $k = 2\pi\ell/d$ are folded back to the center of the Brillouin zone and, thanks to Fourier components introduced by the modulation of either the elastic or photoelastic constants, phonons with $k = 2\pi\ell/d \pm q_S$ ($q_S$ is the scattering wavevector) can become Raman active.[65,66] Here *k* is the magnitude of the wavevector parallel to the growth direction, *d* is the superlattice period and $\ell$ is an integer. These so-called folded acoustic phonons have been extensively studied with spontaneous Raman scattering methods[66] and, more recently, the generation of coherent folded acoustic phonons using light pulses has received much attention.[67,68,69,70] A simple calculation using parameters from the literature[71,72,73] shows that the frequency of the longitudinal-acoustic mode at $k = 2\pi/d$ is close to that of the high-frequency oscillation in Fig. 17. This, as well as the facts that the associated Raman tensor is diagonal, which agrees with the observed selection rules, and that phonon frequencies do not depend on $B$, support our assignment that the oscillation at 11.3 cm$^{-1}$ is due to coherent acoustic phonons.

## IV. CONCLUSIONS

We have presented a comprehensive study, combining spontaneous Raman scattering and coherent time-domain spectroscopy, of low-lying excitations and states associated with the bandgap of a CdTe QW doped with donors and manganese ions. For excitation below the bandgap, our results confirm that there is a system of localized excitons coupled to paramagnetic impurities in a CdTe QW that is well described by the level structure of Fig. 2. We have shown that such a system can be optically excited to generate many-spin Raman coherences and, thus, entanglements involving multiple donors and, independently, Mn$^{2+}$ ions. Our system of paramagnetic impurities is a promising candidate for meeting the five criteria put forth by DiVincenzo for the physical realization of a quantum computer.[74] Explicitly, (*i*) the qubits embodied by the impurity spin states are well characterized and fully scalable, (*ii*) a well-defined initial state can be simply attained by cooling the sample down to sufficiently low temperatures, and (*iii*) spin-flip decoherence times are some of the longest known in the solid phase.[51,75] We further note that the localization centers associated with, say, surface roughness, and the donors need not be at the same sites. Hence, (*iv*) spectral discrimination coupled with submicrometer-sized apertures[10] can possibly be used to excite particular excitons to address a particular set of impurities. Finally, (*v*) our observations of two- and three-qubit entanglement can be construed as a demonstration of light-controlled interaction between the qubits and, as such, they represent a crucial step for the implementation of a universal set of quantum gates.

Other than the results on entanglement mediated by localized excitons, we have shown that photoexcitation of free excitons significantly enhance the degree of nominally-forbidden entanglement of a pair of donor electrons, although the mechanism by which this is achieved could not be identified. We also reported on the observation of exciton Zeeman beats in the Faraday configuration, Raman beats of antiferromagnetically-coupled manganese pairs and folded acoustic phonons.

## ACKNOWLEDGEMENTS

The authors wish the thank Professor A. K. Ramdas for helpful discussions. Work supported by the NSF under Grants No. PHY 0114336 and No. DMR 0245227, by the AFOSR under contract F49620-00-1-0328 through the MURI program and by the DARPA-SpinS program. Acknowledgment is made to the donors of The Petroleum Research Fund, administered by the ACS, for partial support of this research. One of us (AVB) acknowledges partial support from CONICET, Argentina.

**FIGURE CAPTIONS**

FIG. 1. (a) Energy level diagram describing Stokes Raman scattering. (b) Schematic frequency spectrum of the transmitted pump pulse. The Raman coherence between states |1> and |2> is driven by pairs of fields of frequencies ω and ω-Ω, contained within the pulse spectrum.

FIG. 2. Generic level structure of a QW system of bound electrons coupled to a localized exciton of energy $E_e$. Optical transitions are depicted by arrows. **B** is the external magnetic field and $\mathbf{B}_e$ is an effective field describing the interaction between the electrons and the exciton heavy-hole. The energy diagram is not to scale ($E_e$ is much greater than the Zeeman-splitting). Parabolas represent the vibrational analog for $S \gg 1$.

FIG. 3. PL and PLE spectra at $B = 0$ and $T = 2$ K. The polarizations of the incident and scattered light were not analyzed.

FIG. 4. Diagrams showing field-induced splitting of electron and heavy-hole states in the Faraday (left) and Voigt (right) geometry. $E_e$ is the energy of the exciton at zero magnetic field. The dominant PL emission and electric-dipole-allowed transitions for circularly polarized light ($\sigma^+$ and $\sigma^-$) are denoted by arrows.

FIG. 5. (a) PL at $T = 2$ K and various magnetic fields in the Faraday and Voigt geometries. (b) Magnetic field dependence of the PL shift. Lines are fits using Eqs. (15) and (16); see text.

FIG. 6. Raman spectrum in the Voigt configuration for $B = 2.6$ T and $T = 3.5$ K. The laser energy is 1.66 eV. Bars indicate the manganese PR and its overtones. The peak labeled 1SF at ~ 12 cm$^{-1}$ is the spin-flip transition of electrons bound to donors. The magnetic field

dependence of the 1SF frequency is shown in the inset together with the PL shift in the Voigt configuration. Curves are theoretical predictions, Eq. (15), with $x = 0.004$.

FIG. 7. Raman spectrum at $B = 6.89$ T and $T = 5$ K showing $Mn^{2+}$ multiple spin-flip scattering. The broad feature is the PL. The laser energy is 1.685 eV. Inset: Comparison between the PL and the PR resonant Raman excitation (RRS) spectrum (squares).

FIG. 8. Differential magnetic Kerr data at $T = 5$ K. (a) $\hbar\omega_C = 1.60$ eV (below the QW gap) and (b) $\hbar\omega_C = 1.655$ eV (near resonance with localized excitons). Curves are fits using the LP method. Mode parameters from the fits were used to generate the associated Fourier transform spectra shown in the insets. The main feature in (a) is the spin-flip transition of bulk CdTe electrons. In (b), the sharp feature at $\sim 6.5$ cm$^{-1}$ is the $Mn^{2+}$ PR. The peak at $\sim 9$ cm$^{-1}$ is the spin-flip of donor electrons in the QW.

FIG. 9. (a) Pump-induced rotation data for $\hbar\omega_C = 1.682$ eV ($T = 2$K). Curves are fits using the LP-method. Inset: Fourier transform spectrum. The low- (high-) frequency component of the doublet is the spin-flip of donor electrons (electron spin-flip of the exciton). Peak labeled 2SF, at twice the frequency of the donor spin-flip, reflects coherence involving two electrons. The sharp feature is the $Mn^{2+}$ PR which dominates at delays $> 15$ ps. The weak feature at 7 cm$^{-1}$, labeled PR(e), is the excited-state PR. (b) Frequency versus magnetic field. SF and 2SF curves are fits with $x = 0.004$ and $T = 5.5$K.

FIG. 10. Pump-induced rotation data at various magnetic fields. The laser energy and temperature are the same as in Fig. 9. Oscillations above $\sim 15$ ps are due to the $Mn^{2+}$ PR. The Fourier transform spectra of the insets show only the 1SF and 2SF donor transitions.

FIG. 11. Time-domain data at $T = 2$ K and $\hbar\omega_C = 1.687$eV. (a) Differential magnetic Kerr traces at three values of the magnetic field. Only donor-related transitions are shown in the associated Fourier transform spectra. The first (second) overtone of the electron spin-flip is denoted by 2SF (3SF). (b) Frequency vs. magnetic field for the donor spin-flip fundamental, 1SF, and its overtones. Curves are fits with $x = 0.004$ and $T = 5.5$ K.

FIG. 12. Dependence of the intensity of the $Mn^{2+}$ PR on laser energy at $B = 6.89$ T. Comparison between resonant spontaneous Raman (circles; same as in the inset of Fig. 7) and time domain data (squares; the abcisa is $\hbar\omega_C$).

FIG. 13. Differential magnetic Kerr data at $\hbar\omega_C = 1.70$ eV (above the QW gap), $B = 7$ T and $T = 2$K. Inset: Fourier transform spectrum.

FIG. 14. Same as Fig. 13 but for $\hbar\omega_C = 1.71$ eV.

FIG. 15. Differential reflectivity data in the Faraday configuration showing exciton quantum beats ($B = 7$ T, $T = 2$ K and $\hbar\omega_C = 1.67$eV). The inset is the Fourier transform spectrum.

FIG. 16. Magnetic field dependence of the exciton beat frequency (squares) and PL shift in the Faraday configuration (circles; see Fig. 5).

FIG. 17. Differential reflectivity data (after subtracting the exponentially decaying background shown in the top trace) in the Faraday configuration at three values of the magnetic field; $T = 2$K and $\hbar\omega_C = 1.67$eV. Pump and probe beams are linearly polarized. The insets show associated Fourier transform spectra. AFMR denotes the antiferromagnetic resonance of $Mn^{2+}$ pairs.

**FIG. 1**

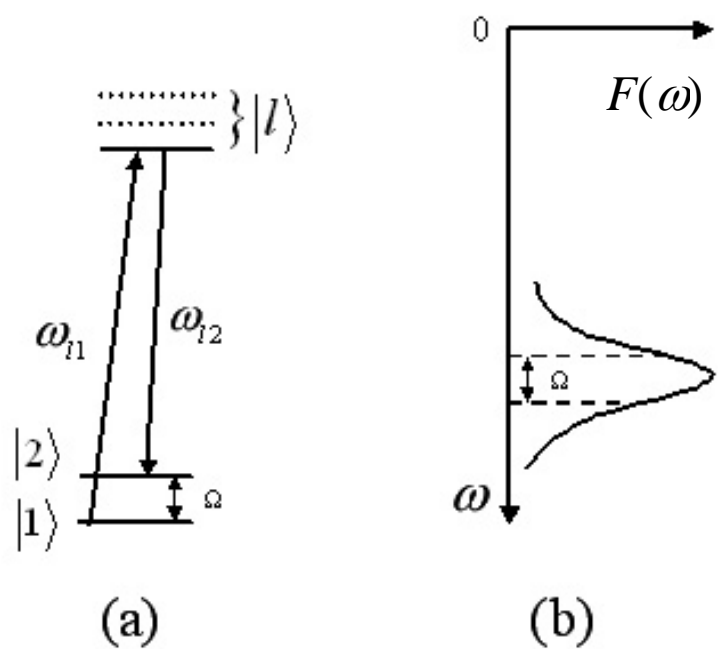

**FIG. 2**

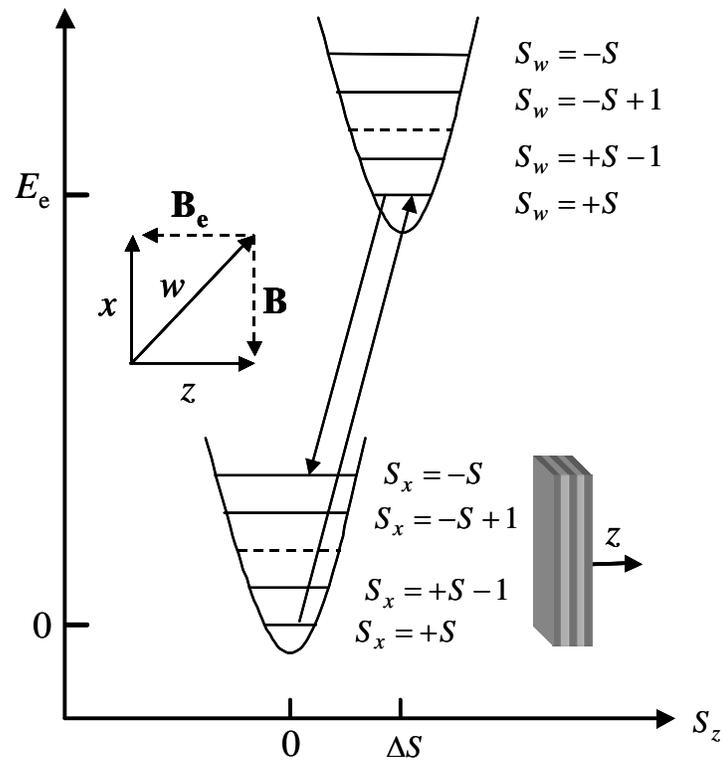

**FIG. 3**

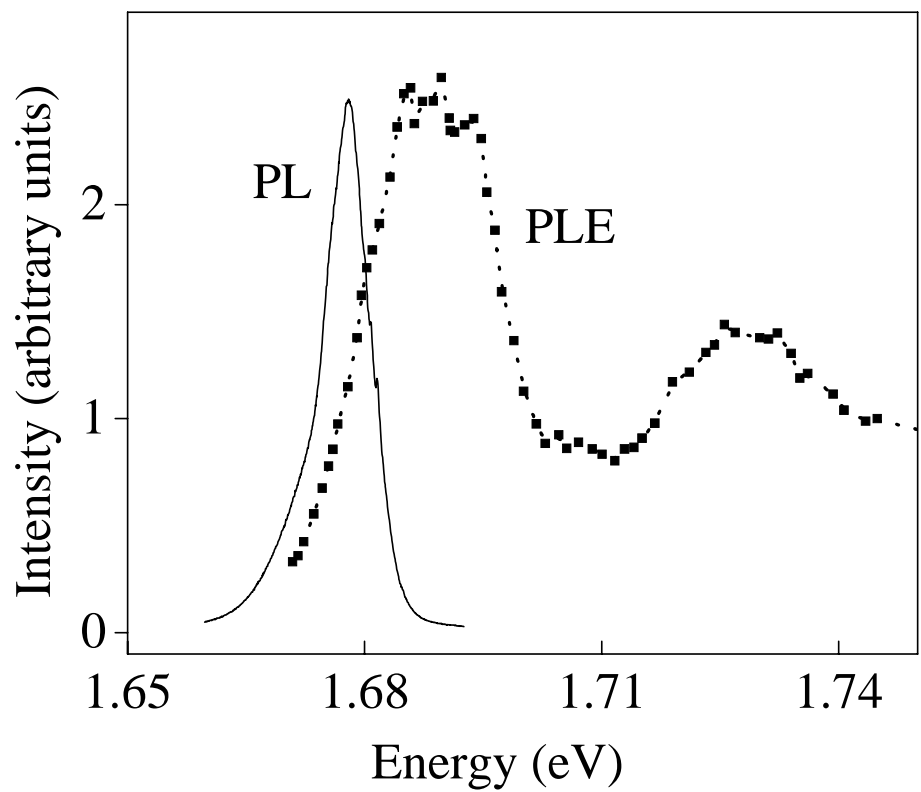

**FIG. 4**

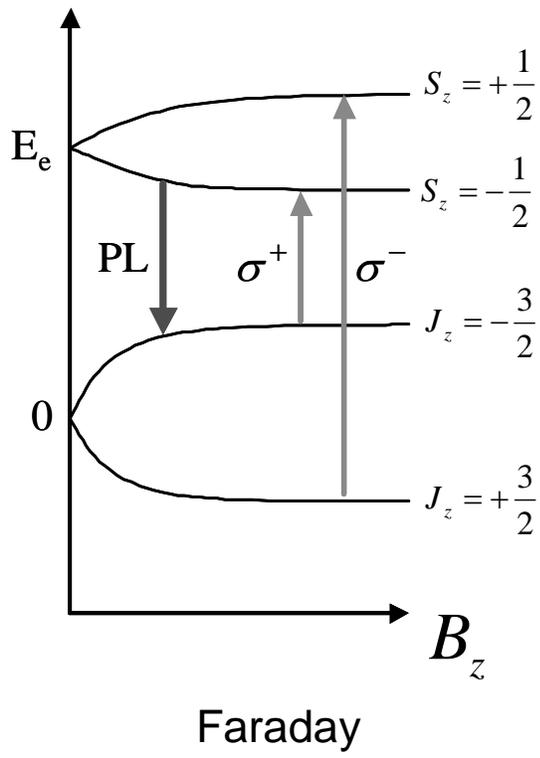
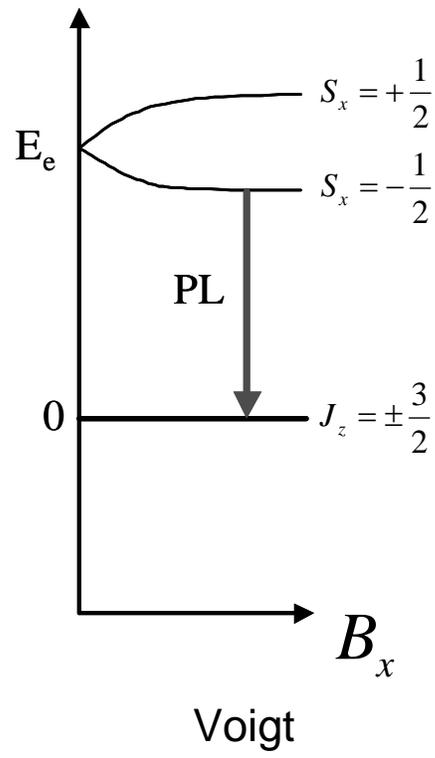

Faraday  Voigt

**FIG. 5**

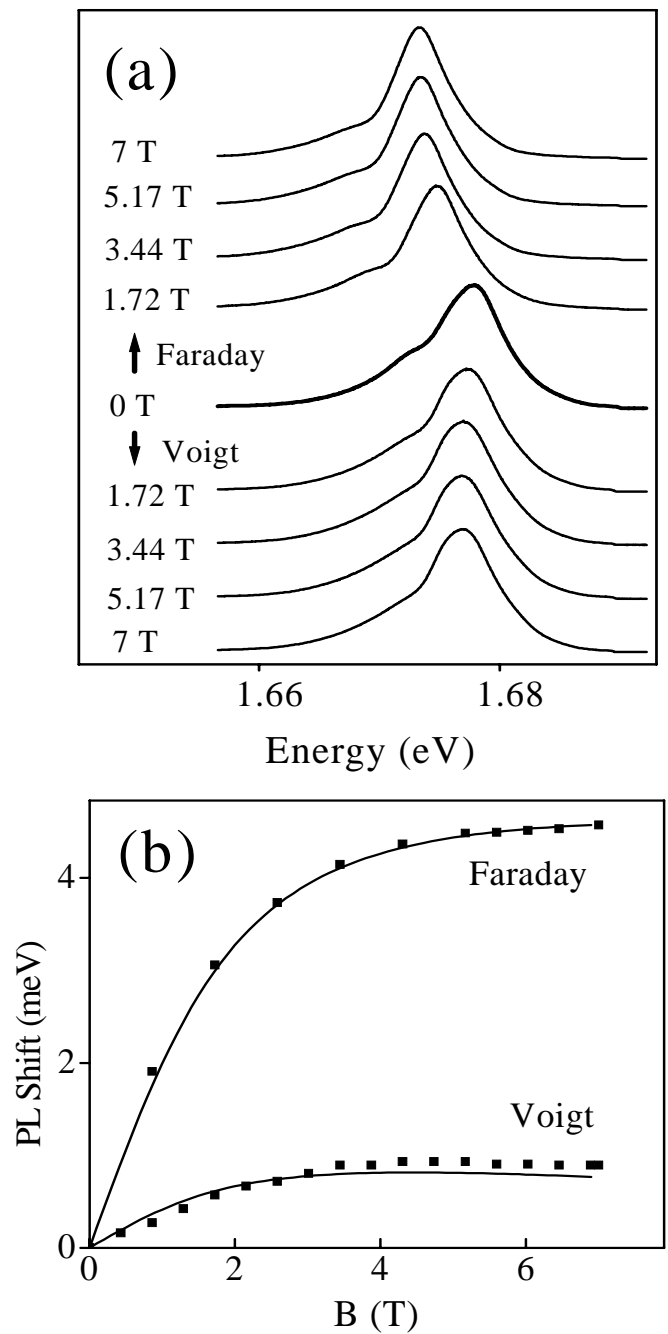

**FIG. 6**

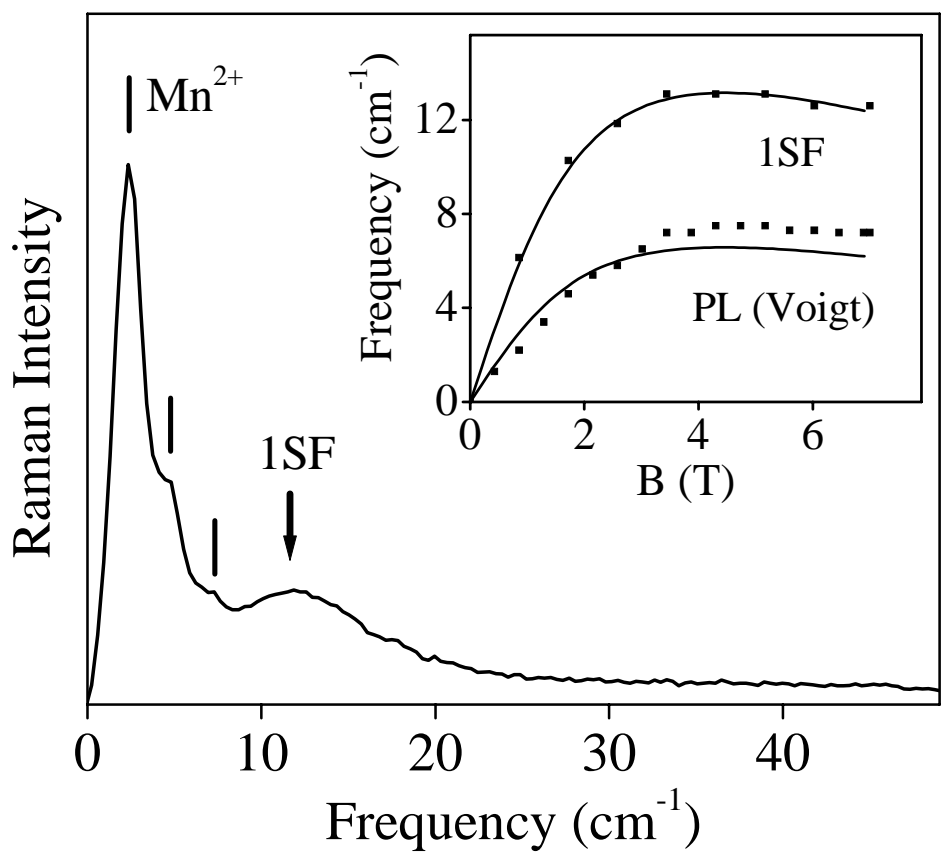

**FIG. 7**

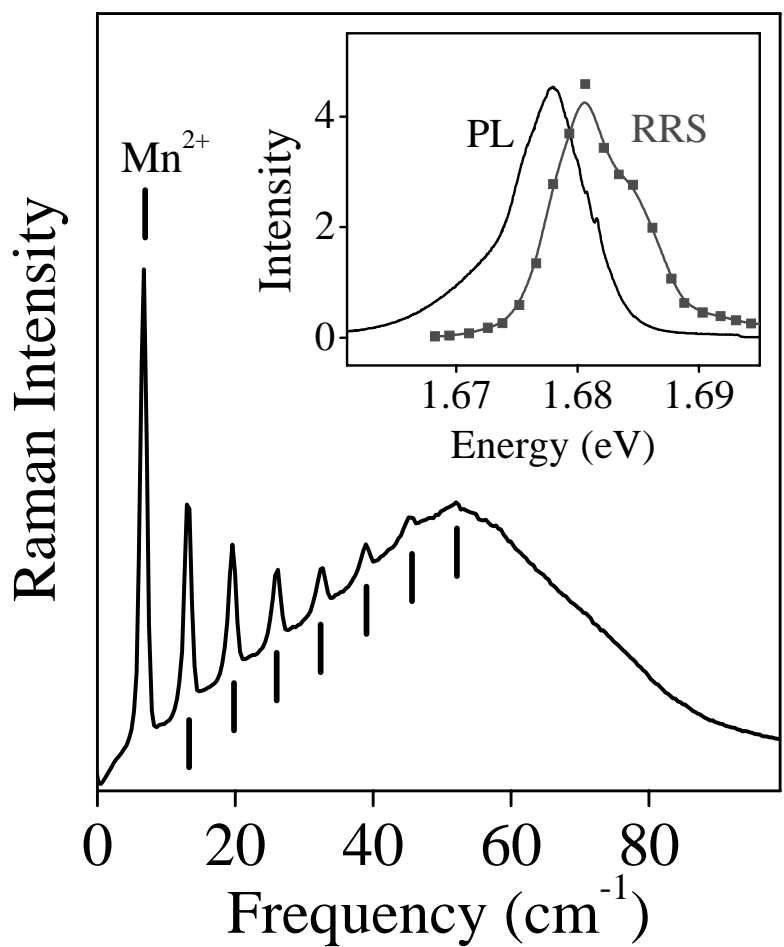

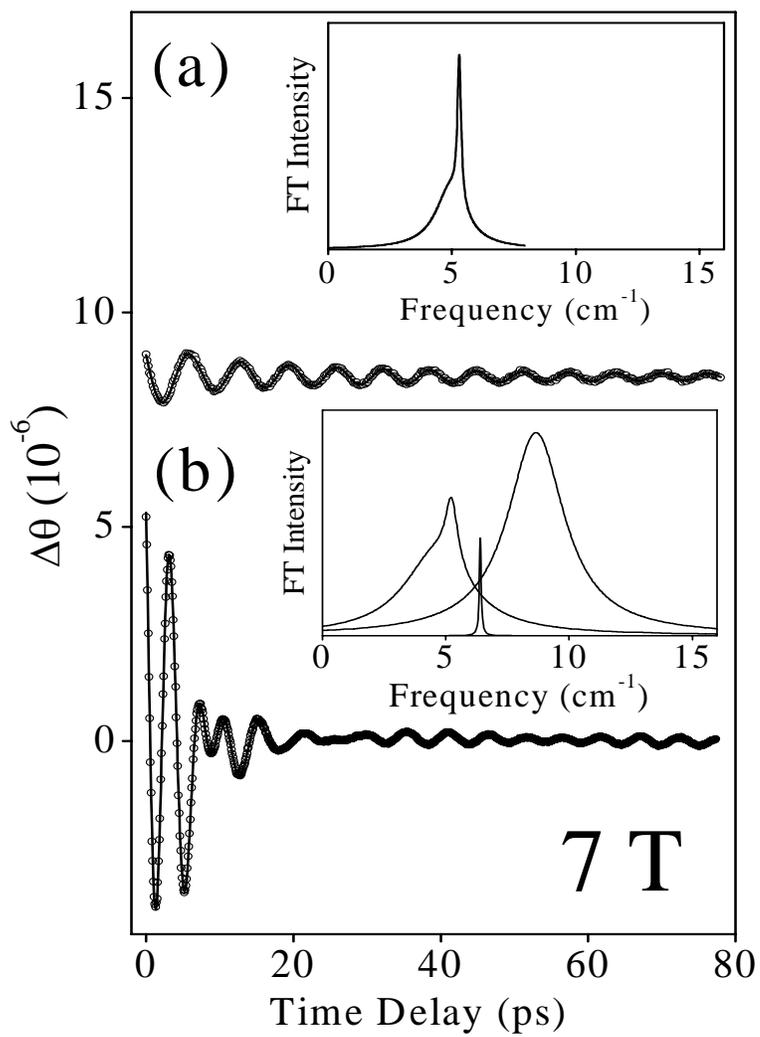

**FIG. 9**

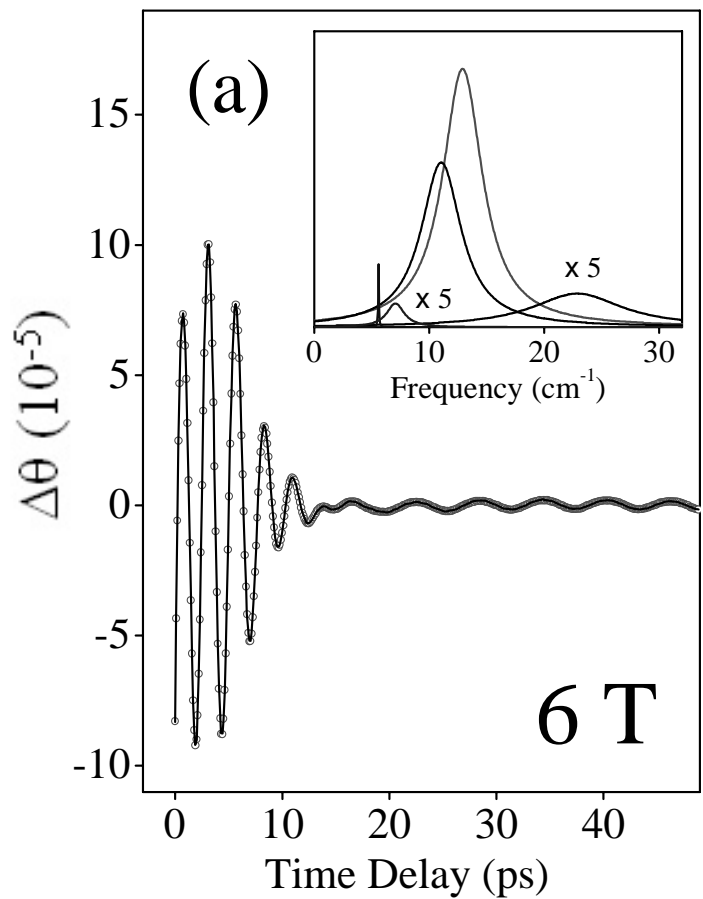

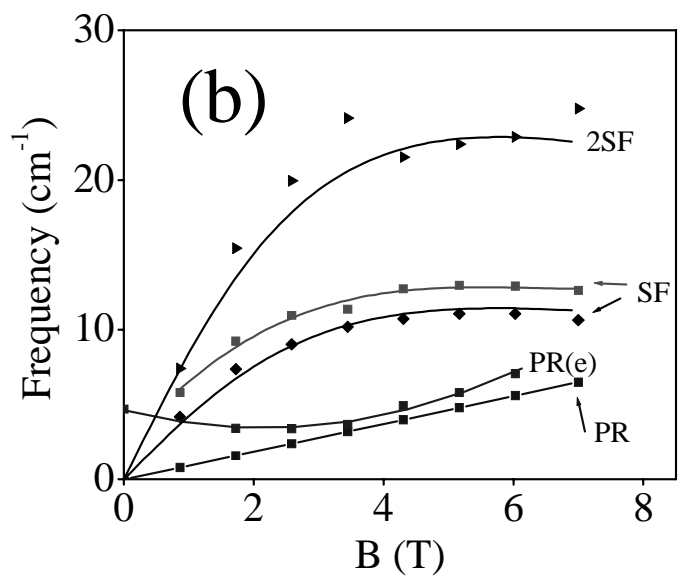

**FIG. 10**

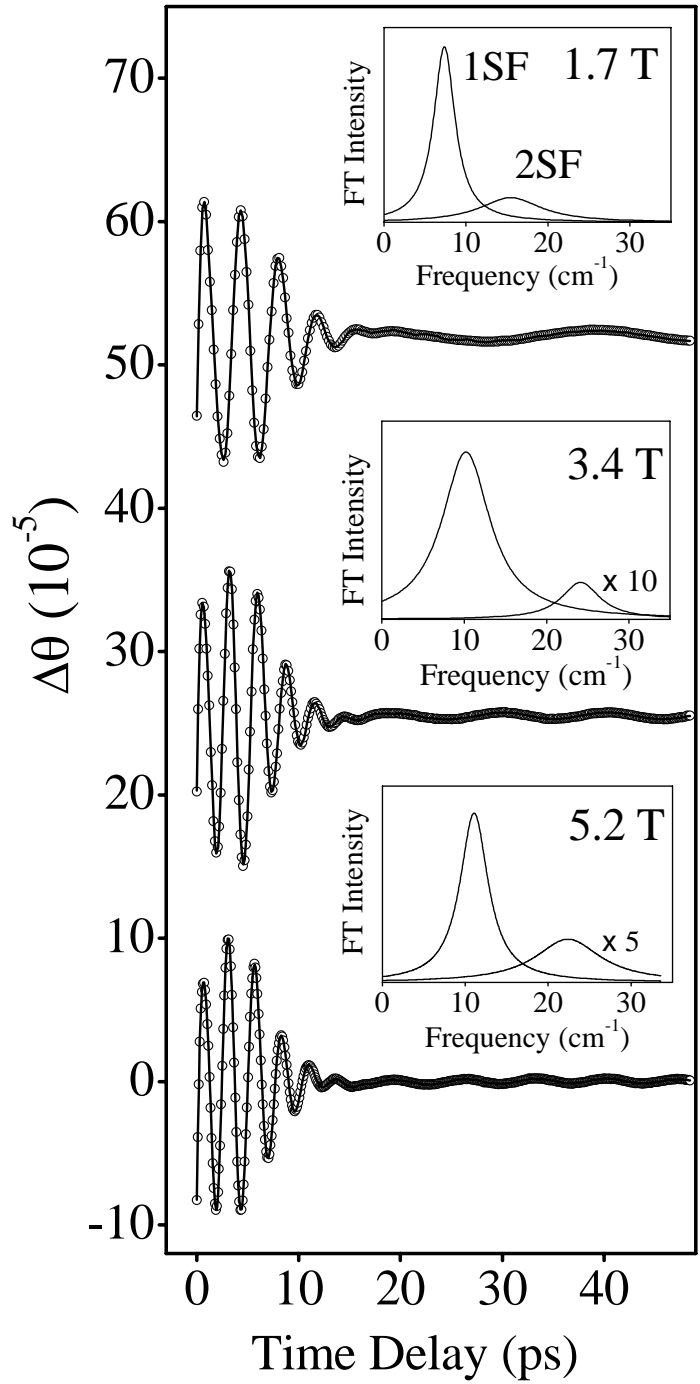

**FIG. 11**

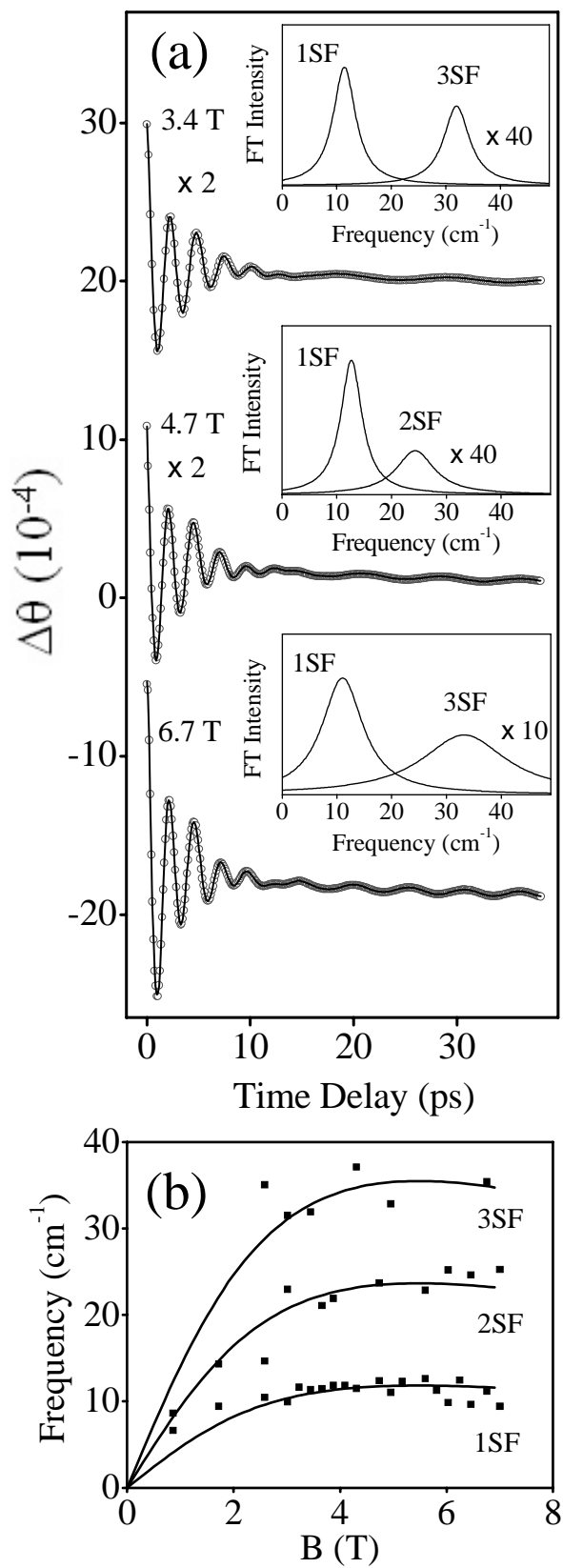

**FIG. 12**

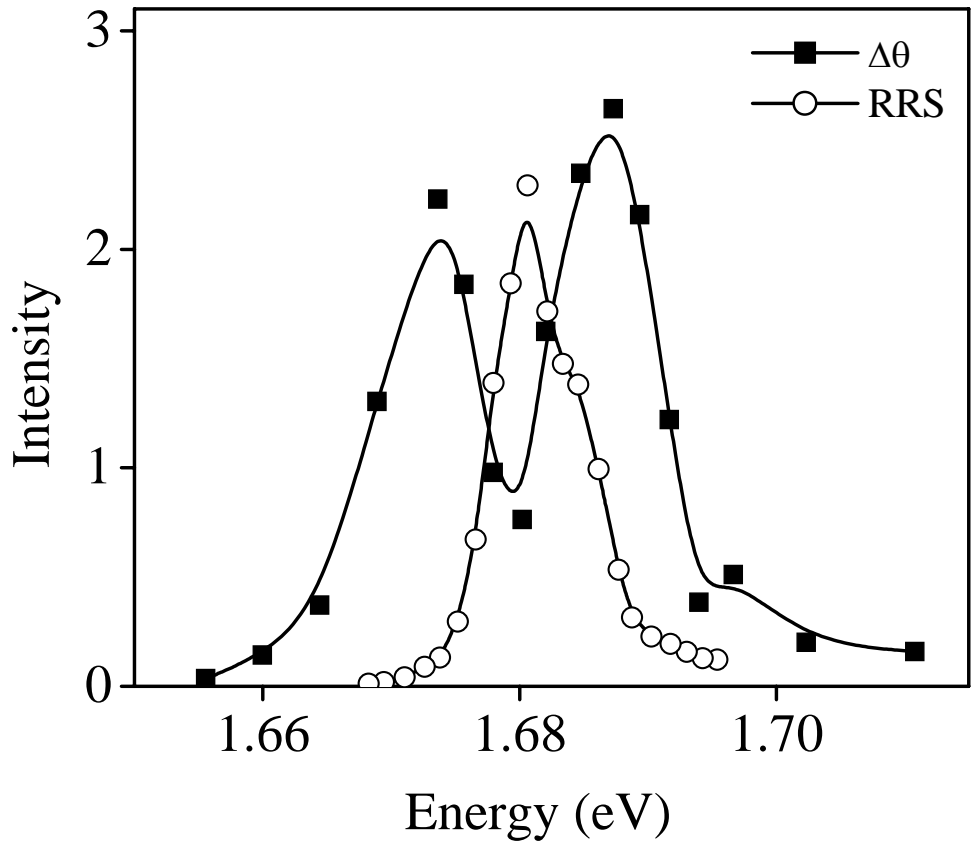

**FIG. 13**

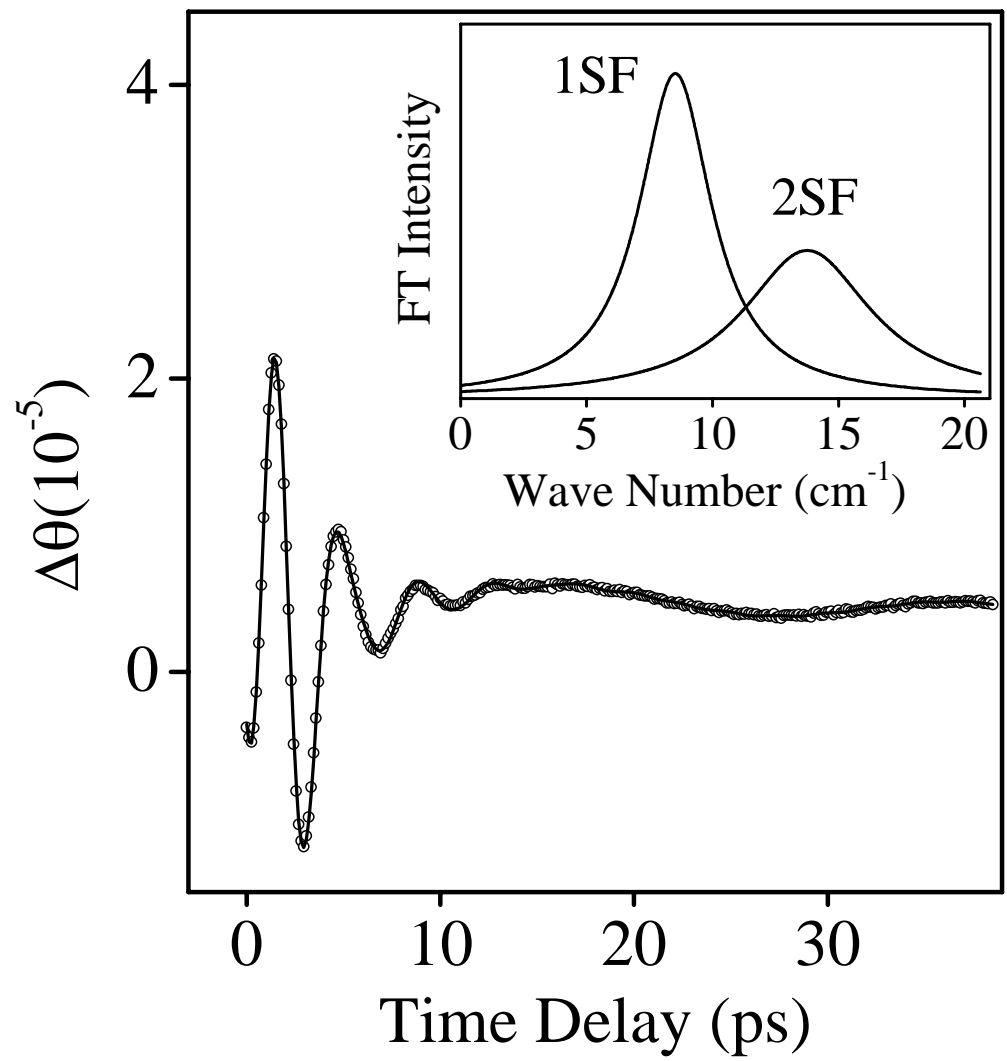

**FIG. 14**

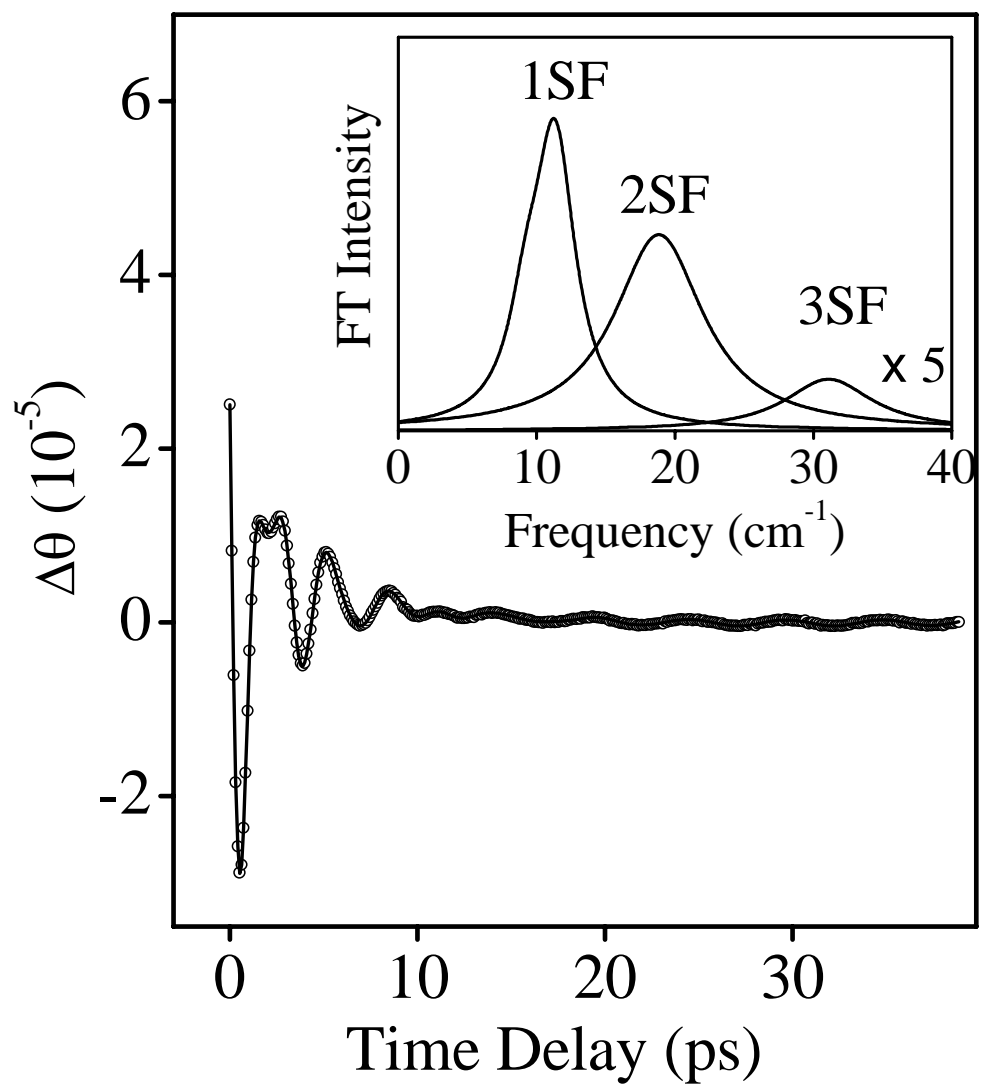

**FIG. 15**

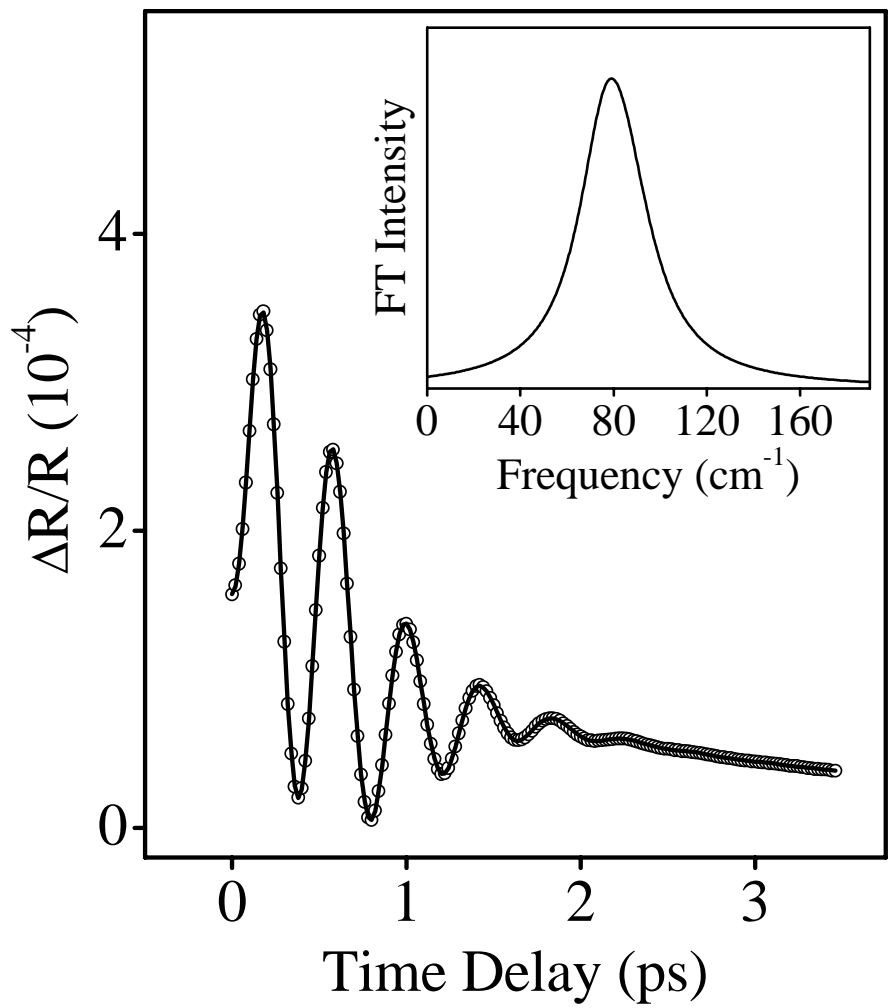

**FIG. 16**

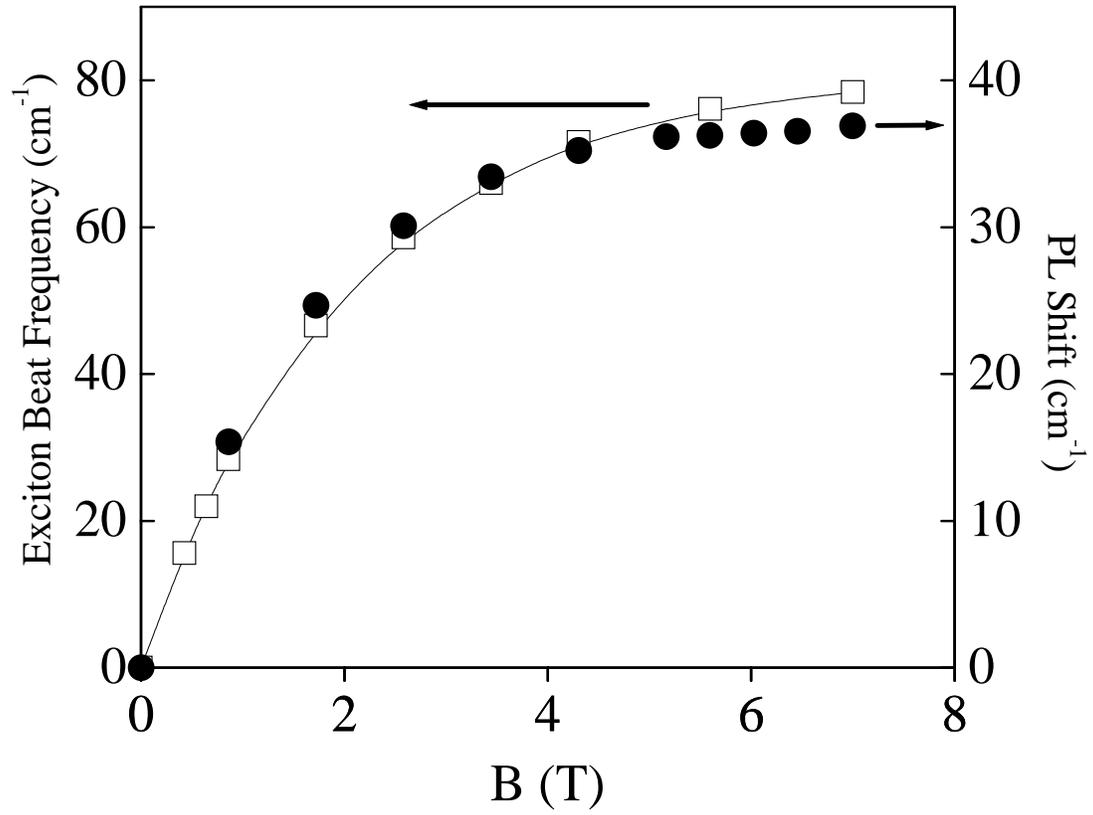

**FIG. 17**

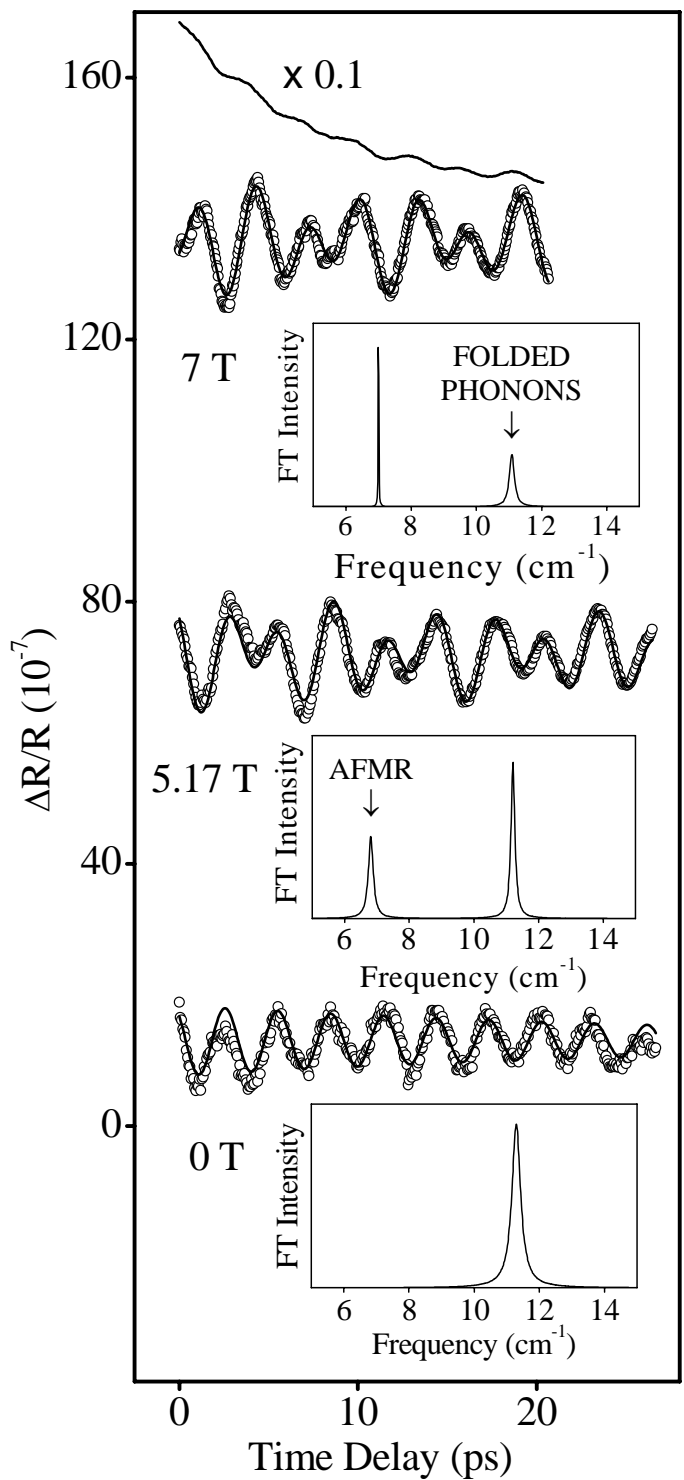